\documentclass[]{aa}
\usepackage{natbib}
\bibpunct{(}{)}{;}{a}{}{,}
\usepackage[pdftex]{graphicx,hyperref}
\usepackage{txfonts}
 \usepackage{subfigure}
\title{The application of compressive sampling to radio astronomy}
\subtitle{\rm{II}: Faraday rotation measure synthesis}

\author{F. Li \inst{1}, S. Brown \inst{1}, T. J. Cornwell \inst{1}, F. de Hoog \inst{1} }
\offprints{F. Li}
\institute{Commonwealth Scientific and Industrial Research Organization (CSIRO), Australia}

\begin{document}

\abstract{ Faraday rotation measure (RM) synthesis is an important tool to study and analyze galactic and extra-galactic magnetic fields. 
Since there is a Fourier relation between the Faraday dispersion function and the polarized radio emission, full reconstruction of the dispersion function requires knowledge of the polarized radio emission at both positive and negative square wavelengths $\lambda^2$. However, one can only make observations for $\lambda^2 > 0$. Furthermore observations are possible only for a limited range of wavelengths. Thus reconstructing the Faraday dispersion function from these limited measurements is ill-conditioned. In this paper, we propose three new reconstruction algorithms for RM synthesis based upon compressive sensing/sampling (CS). These algorithms are designed to be appropriate for Faraday thin sources only, thick sources only, and mixed sources respectively. Both visual and numerical results show that the new RM synthesis methods provide superior reconstructions of both magnitude and phase information than RM-CLEAN. 
}
 
 
\date{Received date; accepted date}
\titlerunning{CS-based RM Synthesis}
\maketitle
\section{Introduction}
\label{sec:intro} 
The intrinsic polarization of a synchrotron emitting source together with knowledge of propagation effects through intervening media provide critical diagnostics for magnetic field orientation and fluctuations in a wide range of astrophysical contexts. Faraday rotation is a physical phenomenon where the position angle of linearly polarized radiation propagating through a magneto-ionic medium is rotated as a function of frequency. As introduced in~\citet{Brentjens:2005p3385} and~\citet{Heald:2009p3423}, Faraday rotation measure synthesis is an important tool for analysing radio polarization data where multiple emitting regions are present along a single line of sight. Observations of extragalactic sources, which by necessity must be viewed through the Faraday rotating and emitting Galactic interstellar-medium~\citep{de2006radio, brown2009diffuse, schnitzeler2007wsrt, schnitzeler2009wsrt}, are an obvious example of this regime. \citet{Burn:1966p3487} introduced the Faraday dispersion function $F(\phi)$, which describes the intrinsic polarized flux per unit Faraday depth $\phi$ (in \rm{rad}~\rm{m}$^{-2}$), and its relationship with the complex polarized emission $P(\lambda^2)$ as
 \begin{equation}
\label{e:faradaytransform}
 P(\lambda^2)=\int_{-\infty}^{\infty} F(\phi)\mathrm{e}^{2\mathrm{i}\phi \lambda^2}\;\mathrm{d}\phi,
\end{equation} where $\lambda$ is the wavelength. Note that $P$ can also be written as $P=Q+\mathrm{i}U$, where $Q$ and $U$ represent the emission of Stokes $Q$ and Stokes $U$, respectively. 

To study multiple emitting and Faraday rotating regions along each line of sight, we need to reconstruct the Faraday dispersion function, which is, in general, a complex-valued function of the Faraday depth $\phi$. From Eq. (\ref{e:faradaytransform}) , we can invert the expression to yield:
 \begin{equation}
\label{ e:invfaradaytransform}
 F(\phi)=\frac{1}{\pi}\int_{-\infty}^{\infty} P(\lambda^2)\mathrm{e}^{-2\mathrm{i}\phi \lambda^2}\; \mathrm{d}\lambda^2.
\end{equation} 
However, the problem is that we can not observe the polarized emission at wavelengths where $\lambda^2<0$. Even for the wavelength range $\lambda^2>0$, it is impossible to observe all wavelengths or frequencies. \cite{Brentjens:2005p3385} propose a synthesis method by first introducing an observing window function $M(\lambda^2)$. The observed complex polarized  emission can then be described as
 \begin{equation}
\label{e:observedflux}
 \widetilde{P}(\lambda^2)=M(\lambda^2){P}(\lambda^2).
  \end{equation} In this paper, the tilde denotes the observed quantities. 

If the observing window function is $M(\lambda^2)$ with $m$ channels, the RM spread function (RMSF) is be defined by
\begin{equation}
\label{e:rmsf}
R(\phi)= K \sum_{i=1}^{m}M(\lambda_i^2)\mathrm{e}^{-2\mathrm{i}\phi( \lambda^2_i-\lambda^2_0)},
\end{equation}  where the parameter $\lambda^2_0$ is the mean of the sampled values between $\lambda^2_1$ and  $\lambda^2_m$ within the observation window $M(\lambda^2)$;  $i$ is the $i^{\rm{th}}$ channel in the observation window, and $K$ is a normalising constant of the window function $M(\lambda^2)$. In this paper, we assume as a simplification that all channels have uniform weights for the $m$ channels in the observing window function.

In~\citet{Brentjens:2005p3385}, the reconstructed Faraday rotation measure synthesis can be written in discrete form as
\begin{equation}
\label{e:discrete_invfaradaytransform}
 \widetilde{F}(\phi) \approx K \sum_{i=1}^{m}\widetilde{P}(\lambda^2_i)\mathrm{e}^{-2\mathrm{i}\phi( \lambda^2_i-\lambda^2_0)}, 
\end{equation} where  $\widetilde{F}(\phi)$ is the reconstructed Faraday dispersion function. From Eq. (\ref{e:discrete_invfaradaytransform}), we can see that the Faraday dispersion function can be  reconstructed provided that the spectral coverage is sufficient. 

However, the reconstructed results generally include some side lobes. Using the terminology of radio
interferometry, the result of Brentjens \& de Bruyn' method is a dirty version of the Faraday dispersion function and is abbreviated as ``the dirty curve". It is the convolution of $F(\phi)$ and the RMSF, and a deconvolution step may be used to clean it up. By borrowing the cleaning procedure in the image deconvolution method of H\"ogbom 
CLEAN~\citep{Hogbom:1974p1206}. ~\cite{Heald:2009p3423} proposes the RM-CLEAN method which deconvolves $\widetilde{F}(\phi)$ with the RMSF to remove the sidelobe response. 

Recently, \cite{Frick:2010p3495} proposed a wavelet-based Faraday RM synthesis method. In that approach, the authors assume specific magnetic field symmetries in order to project the observed polarization emissions onto $\lambda^2<0$. 
 
%

Compressive sensing/sampling (CS)~\citep{Candes:2008p14, Candes:2006p23,Wakin:2008p1623} has been one of the most active areas in signal and image processing over the last few years. Since CS was proposed, it has attracted very substantial interest, and has been applied in many research areas~\citep{Wakin:2006p1437,Lustig:2007p1719,Puy:2010p1807, Mishali:2009p2147,Bobin:2009p2006}. In radio astronomy, CS has attracted attention as a tool for image deconvolution. \citet{Wiaux:2009p2267} compare the CS-based deconvolution methods with the H{\"o}gbom CLEAN method~\citep{Hogbom:1974p1206} on simulated uniform random sensing matrices with different coverage rates. They apply compressive sampling for deconvolution by assuming the target signal is sparse.~\cite{Wiaux:2009p1697}  proposed a new spread spectrum technique for radio 
interferometry by using the non-negligible and constant component of the 
antenna separation in the pointing direction. Recently, a new CS-based image deconvolution method was introduced in~\cite{deconvolution2011application} in which an isotropic undecimated wavelet transform is adopted as a dictionary for sparse representation for sky images. 

In this paper, we propose three new CS-based RM synthesis methods. In Section~\ref{sec:csrm}, the three CS-based 
RM synthesis methods are proposed. The implementation details of the general experiment layout is given in Section~\ref{sec:implementation}. Simulation results from the traditional methods are 
compared with those from CS-based methods in Section~\ref{sec:experiments}. 
The final conclusions are given in Section~\ref{sec:conclusion}.	

%
%

\section{CS-based RM synthesis}
\label{sec:csrm} 

CS is primarily a sampling theory for sparse signals. A sensing matrix~\citep{candes2006stable} is used to sample a signal with sparsity (few non-zero terms) or a sparse representation with respect to a given basis function dictionary. Given a limited number of measurements, generally less than the number of unknowns in the target signal, the target signal can be reconstructed by optimisation of an L1 norm. More information on the key concepts (such as sparsity, incoherence, the restricted isometry property, and the L1 norm reconstruction) and results can be found in~\cite{Candes:2007p2815,Candes:2004p2832,Candes:2008p14,Candes:2006p23}. 

CS includes two steps: sensing/sampling and reconstruction. This is in contrast to Nyquist-Shannon theory which measures the target signal directly without the reconstruction step. In this paper, we will focus on the reconstruction step (calculating the Faraday dispersion function given an observing window) rather than the sensing step (the selection of the observing window), because the observing frequency range and the bandwidth for each channel are usually fixed for a given telescope array.
 

To proceed with the CS approach, we rewrite the Fourier relationship as a matrix equation. The projection of the Faraday dispersion function to the polarized emission can be described as a matrix $Y$ of size $m\times N$ 
\begin{equation}
\label{Faradaytrans}
\mathrm{Y}(j,N/2+k)=\mathrm{e}^{2\mathrm{i}\phi_k\lambda^2_j}, j={1},\cdots,{m}  ;  k={1-N/2},\cdots,{N/2}. 
\end{equation} The inverse of the projection is the conjugate transpose of $Y$
\begin{equation}
\label{Faradaytrans}
\mathrm{Y}^\ast(N/2+k,j)=\mathrm{e}^{-2\mathrm{i}\phi_k\lambda^2_j}, j={1},\cdots,{m}  ;  k={1-N/2},\cdots,{N/2},
\end{equation} where $\ast$ denotes the conjugate transpose. Suppose $f$ denotes the original Faraday dispersion function $F(\phi)$ in a vector format of length $N$, then the relationship between the Faraday dispersion function and the observed radio emission is:
\begin{equation}
\label{explainRM}
\mathrm{Y}\mathbf{f}=\widetilde{\mathbf{p}},
\end{equation} where $\widetilde{p}$ denotes the observed polarized emission in a vector format of length $m$.

Because we can only measure a limited number of observations with the limited number of channels, i.e. $m<<N$, there are many different potential Faraday dispersion functions consistent with the measurements. To resolve these ambiguities, the usual approach is to use some prior information to select a solution. The prior information can be: the Faraday dispersion function is real; the Faraday dispersion function has only point like signals which are sparse in the Faraday depth domain or the Faraday dispersion function has a sparse presentation with respect to a dictionary of basis functions, to name just a few. Our three synthesis methods are based upon the last two structural assumptions. 

Before introducing our new RM synthesis methods, we need to review two technical terms: Faraday thin and Faraday thick. A source can be either Faraday thin if $\lambda^2\bigtriangleup \phi\ll 1$, or Faraday thick if $\lambda^2\bigtriangleup \phi\gg1$, where $\bigtriangleup \phi$ is the extent of the source along the axis of Faraday depth $\phi$. Faraday thin sources can be well described by Dirac $\delta$ function of $\phi$, while Faraday thick sources have extensive support on the Faraday depth axis~\citep{Brentjens:2005p3385}. Note that the definition of Faraday thin or thick is wavelength dependent. 

\subsection{RM synthesis for Faraday thin sources}
\label{subsec:thin} 

The relationship between the Faraday dispersion function and the observed polarized radio emission is a Fourier pair if $\lambda^2=\pi u$, where $u$ is a wavelength related parameter. Since the space and Fourier domain are perfectly incoherent~\citep{Candes:2007p2815}, we can apply CS for RM synthesis in a straightforward manner provided there are Faraday thin sources along the line of sight since the screen is necessarily sparse.   

In this context, CS recommends solving for the Faraday dispersion function by minimising the L1 norm (summed absolute value) of the dispersion function as inimising the L1 norm optimises the sparsity of the reconstruction. There remains one further obstacle - the dispersion function is complex. We handle this by summing the L1 norm of the real and imaginary parts:

\begin{equation}
\label{e:faraday_thin}
{\rm{min}} \;\{ {\| \mathrm{Re}(\mathbf{f})\|_{l_1} + \|\mathrm{Im}(\mathbf{f})\|_{l_1}}  \} \; \;\; s.t.\; \mathrm{Y}\mathbf{f}=\widetilde{\mathbf{p}},
\end{equation} where Re$(\bullet)$ and Im$(\bullet)$ denote the real and the imaginary parts, respectively. By forming a real-valued vector of double length (comprising of the real part and the imaginary part) of the complex-valued vector, almost all L1 norm optimization solvers can be used for Eq.~\ref{e:faraday_thin}. This CS-based rotation measure synthesis for Faraday thin sources is abbreviated as CS-RM-Thin. This is similar in concept to RM-CLEAN because the assumption for RM-CLEAN is that the Faraday dispersion function comprising of spike like signals. However, results in Section~\ref{sec:experiments} show that CS-RM-Thin can provides superior results to RM-CLEAN.

\subsection{RM synthesis for Faraday thick sources}
\label{subsec:thick}

CS-RM-Thin can work effectively when the Faraday dispersion function includes Faraday thin sources along the line of sight. This limits its application for the case when
there are some Faraday thick sources along the line of sight. However, CS can still reconstruct the Faraday dispersion function efficiently provided that we can find a suitable dictionary of basis functions that can decompose the 
extended sources into a sparse representation as described in \citet{Candes:2006p23}. In this paper, we adopt the Daubechies D8 wavelet transforms~\citep{daubechies1992tlw} as the dictionary. Other wavelet transforms can also be adopted, the selection depends on the property of the Faraday dispersion function. We choose the D8 wavelet transform, because we assume that the Faraday dispersion function with thick sources is a sinc-like signal.

We can rewrite Eq.~(\ref{explainRM}) as
\begin{equation}
\label{explainRMwavelet}
\mathrm{Y}\mathrm{W}^{-1}\alpha=\widetilde{\mathbf{p}},
\end{equation}where $\mathrm{W}^{-1}$ is the inverse wavelet transform matrix of size $N \times N$; $\alpha$ is the wavelet coefficient of the Faraday dispersion function $\mathbf{f}$. The wavelet transform matrix is denoted as $\mathrm{W}$, therefore, $\alpha=\mathrm{W}\mathbf{f}$. Other symbols follow the definitions in Eq.~(\ref{explainRM}). Under the condition that $\mathrm{Y}\mathbf{f}=\widetilde{\mathbf{p}} $, we adopt the following assumption: both the real part and the imaginary part of the Faraday thick sources will have a sparse representation in the wavelet domain independently. Then the wavelet based CS RM synthesis method for Faraday thick sources can be written as

\begin{equation}
\label{e:faraday_thick}
{\rm{min}} \;\{ {\| \mathrm{W}\cdot{\rm{Re}}(\mathbf{f})\|_{l_1}+\|\mathrm{W}\cdot{\rm{Im}}(\mathbf{f})\|_{l_1}}  \} \; \;\; s.t.\;  \mathrm{Y}\mathbf{f}=\widetilde{\mathbf{p}}.
\end{equation} This CS-based rotation measure synthesis for Faraday thick sources is abbreviated as CS-RM-Thick.

\subsection{RM synthesis for Faraday mixed sources}
\label{subsec:mixed} 
So far, we have proposed two RM synthesis methods: CS-RM-Thin and CS-RM-Thick for solving Faraday thin sources and thick sources, respectively. However, this  begs the question: which method should be selected if there are both Faraday thin sources and thick sources along the line of sight? Moreover, how can we make a selection if we have no prior information about the Faraday dispersion function, i.e. we are not sure what it looks like? Clearly, neither of them is suitable, we therefore need another solution for solving the above problems. Let us assume that there are both Faraday thin sources and Faraday thick sources in $F(\phi)$. Suppose $\mathbf{f}_{\rm{thin}}$ denotes the Faraday thin sources in $F(\phi)$ in a vector format of length $N$; $\mathbf{f}_{\rm{thick}}$ denotes the Faraday thick sources in $F(\phi)$ in a vector format of length $N$, then $\mathbf{f}_{\rm{thin}}+\mathbf{f}_{\rm{thick}}=\mathbf{f}$. Eq.(\ref{explainRM}) can be rewritten as
\begin{equation}
\label{e:RMseperation}
\mathrm{Y}\mathbf{f}_{\rm{thin}} +  \mathrm{Y}\mathbf{f}_{\rm{thick}}=\widetilde{\mathbf{p}}.
\end{equation} Since we know that the L1 norm can preserve sparsity; Faraday thin sources show sparsity in the Faraday depth domain; Faraday thick sources show sparsity in the wavelet domain, we propose the following solution for the mixed circumstance

\begin{eqnarray}\label{e:faraday_mix}
&{\rm{min}} \;\{     {\| \mathrm{Re}(\mathbf{f}_{\rm{thin}})\|_{l_1}+\| \mathrm{Im}(\mathbf{f}_{\rm{thin}}) } \|_{l_1} +  \|{\mathrm{W}\cdot{\rm{Re}}(\mathbf{f}_{\rm{thick}}) \|_{l_1}}  \nonumber \\
 &  +\|\mathrm{W}\cdot{\rm{Im}}(\mathbf{f}_{\rm{thick}})\|_{l_1} \}\; \;\; \;\; s.t.\;  \mathrm{Y}\mathbf{f}=\widetilde{\mathbf{p}},
\end{eqnarray} where the definition of $\mathrm{W}$ is the same as the above subsection. The above solution for Faraday mixed sources is still based on the spirit of CS by preserving the sparsity in the Faraday depth domain for Faraday thin sources and in the wavelet domain for Faraday thick sources, simultaneously.
This CS-based rotation measure synthesis for Faraday mixed sources is abbreviated as CS-RM-Mix.

\section{Implementation details}
\label{sec:implementation} 
In this section, the implementation details of the above three proposed CS-based RM synthesis methods will be given. 
 \subsection{Preparation}
\begin{figure}
 \begin{center}
     \includegraphics[scale=0.42]{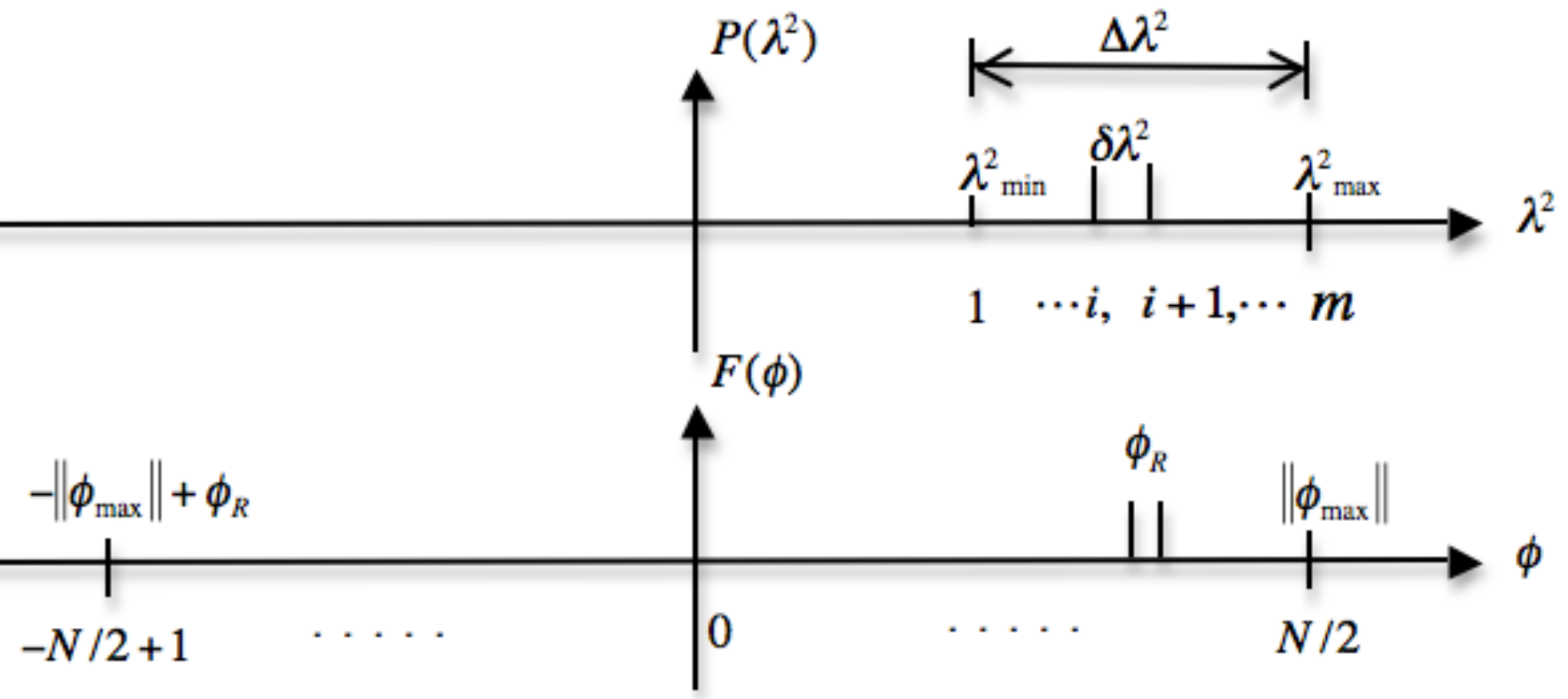}
 \end{center}
\caption{This diagram shows the relationship between parameters in $\lambda^2$ domain and $\phi$ domain, respectively. } \label{fig_wavelength}
\end{figure}

To create a general experiment layout, we borrow some definitions and conclusions from~\cite{Brentjens:2005p3385}. Both the diagram of the wavelength square $\lambda^2$ and $\phi$ are displayed in figure~\ref{fig_wavelength}. The maximum observable Faraday depth $\| \phi_{\mathrm{max}}\|$ is given in~\cite{Brentjens:2005p3385}
\begin{equation}
\label{phimax}
\| \phi_{\mathrm{max}}\|\approx \frac{\sqrt{3}}{\delta \lambda^2},
\end{equation}where $\delta \lambda^2$ is the width of an observing channel. The full width at half maximum (FWHM) of the main peak of the RMSF can be estimated by \begin{equation}
\label{phiresolution}
\delta \phi \approx \frac{2\sqrt{3}}{\bigtriangleup\lambda^2},
\end{equation} where $\bigtriangleup\lambda^2$ is the width of the total $\lambda^2$ distribution.

Before using CS-based RM synthesis, the following two steps are needed:
\begin{enumerate}
\item Select the resolution of the Faraday depth $\phi_R$. Since we know the maximum observable Faraday depth $\phi_{\rm{max}}$ from Eq.~(\ref{phimax}) and the FWHM of the main peak of the rotation measure spread function from Eq.~(\ref{phiresolution}), we can select a grid resolution parameter $\phi_R$ in $\phi$ space, which should be four or five times less than $\delta \phi$, to achieve Nyquist sampling. However, for some observational window functions $M(\lambda^{2})$, the maximum scale (Faraday thickness) that one is sensitive to, estimated by $\frac{\pi}{\lambda^{2}_{\mathrm{min}}}$, is actually smaller than $\delta \phi$. In these cases, it might be practical to Nyquist sample this smaller scale in order to calculate $\phi_{R}$. Based on  $\phi_{\rm{max}}$ and $\bigtriangleup \phi$, we can calculate the number of grid points $N$ as
\begin{equation}
\label{gridnum}
N=\rm{floor}(\frac{2\phi_{\rm{max}}}{\phi_R}).
\end{equation}

\item Constructing the two matrices $\mathrm{Y}$ and $\mathrm{Y}^\ast$.

\end{enumerate}

The selection of these CS-based RM synthesis methods depends on the prior knowledge about the Faraday dispersion function. If we assume it includes Faraday thin sources only along the line of sight, we should select CS-RM-Thin. On the other hand, CS-RM-Thick should be used if we know that there are Faraday thick sources only. When we know that  there are both Faraday thin sources and thick sources along the line of sight, CS-RM-Mix should be used. In most circumstances, we have no prior information about the Faraday dispersion function, CS-RM-Mix can always be used to reconstruct a reliable result as a compromise.

 \subsection{L1 norm solvers for CS-Based RM synthesis methods}
 For CS-RM-Thin and CS-RM-Thick, many optimization methods~\citep{Beck:2009p710,Becker:2009p2860,Boyd:2004p3016} can be used to solve the L1 norm minimization problem in a straightforward manner. There are many
solvers or toolboxes, for example, L1-Magic Matlab toolbox which can be download from~\url {http://www.acm.caltech.edu/l1magic/}.  In this paper, L1-Magic is adopted for solving equations~(\ref{e:faraday_thin}) and (\ref{e:faraday_thick}). Fast Iterative Shrinkage-Thresholding Algorithm (FISTA)~\citep{Beck:2009p710} can also be used for solving this problem if we rewrite Eq.~(\ref{e:faraday_thin}) or~(\ref{e:faraday_thick}) in a Lagrangian form. 

As far as CS-RM-Mix is concerned, the solvers or toolboxes introduced above can be used for solving Eq.~(\ref{e:faraday_mix}) indirectly. Suppose $\alpha_\mathrm{thick}$ denotes the wavelet coefficients of the thick sources $\mathbf{f_\mathrm{thick}}$ in the Faraday dispersion function in a vector format, and $\mathrm{W}^{-1}$ is the inverse wavelet transform matrix, then we have 
\begin{equation}
\label{e:thicksources_express}
\mathbf{f_\mathrm{{thick}}}=\mathrm{W}^{-1}\alpha_\mathrm{thick}.
\end{equation} Bring the above equation into Eq.~(\ref{e:RMseperation}), we have

\begin{equation}
\label{e:derivation_seperation}
\mathrm{Y}\mathbf{f}_{\rm{thin}} +  \mathrm{Y} \mathrm{W}^{-1}\alpha_\mathrm{thick}=\widetilde{\mathbf{p}}.
\end{equation} Furthermore, Eq.~(\ref{e:derivation_seperation}) can be rewritten as
\begin{equation}\label{e:derivation_seperation2}
    \left[\mathrm{Y} \;  \mathrm{Y} \right]
      \left[\begin{array}{cc}
       \mathrm{I} &   \mathrm{O} \\
         \mathrm{O} &  \mathrm{W}^{-1}
           \end{array}
    \right]  \left[\begin{array}{c}
       \mathbf{f}_\mathrm{thin}     \\
         \alpha_\mathrm{thick} 
           \end{array}
    \right]=\widetilde{\mathbf{p}}_{m\times 1}, 
\end{equation} where $\mathrm{I}$ denotes the identity matrix of size $N\times N$, and $\mathrm{O}$ is the matrix of all zeros with the size of $N\times N$.  
If we denotes $\mathrm{Y_{mix}}=[\mathrm{Y}\;\mathrm{Y}]_{m\times 2N}$, $\mathrm{T}$=$ \left[\begin{array}{cc}
       \mathrm{I} &   \mathrm{O} \\
         \mathrm{O} &  \mathrm{W}^{-1}
           \end{array}
    \right]_{2N\times 2N}  $ and $\mathbf{c}=\left[\begin{array}{c}
       \mathbf{f}_\mathrm{thin}     \\
          \alpha_\mathrm{thick}     \end{array}
    \right]_{2N\times 1}$, almost all L1 norm minimization solvers can be used to solve Eq.~(\ref{e:faraday_mix}) with:

 \begin{eqnarray}\label{e:faraday_mix_simple}
\rm{min}} \;\{     {\| \mathrm{Re}(\mathbf{c})\|_{l_1}+\| \mathrm{Im}(\mathbf{c}) \|_{l_1}} \}   \; \;\;  s.t.\;  \mathrm{Y_{mix}T}\mathbf{c}=\widetilde{\mathbf{p}.
\end{eqnarray}


 To help readers who are unfamiliar with L1 norm minimization to use or implement our proposed CS-based RM synthesis methods, we have developed a simple algorithm based on the iterative soft-thresholding algorithm (ISTA)~\citep{Beck:2009p710} for CS-RM-Mix as an example. The algorithm is as follows:


\begin{enumerate}
\item Initialization:
              \begin{enumerate}
                    \item Choose parameters: the soft-threshold $\tau$ (this can be set by 1 for most circumstances) , the stopping-threshold $\delta$ (this can be set by the noise level) 
                    \item Set the number of iteration $l=\rm{floor}(\tau/\delta)$
                     \item $ \mathbf{f}_{\rm{thin}}=\mathrm{Y}^\ast \widetilde{\mathbf{p}}$; $\mathbf{f}_{\rm{thick}}=0$
               \end{enumerate}
 
\item Within $l$ iterations:
              
             \begin{enumerate}
                      \item    Reconstructing the Faraday thin sources

	                            \begin{enumerate}\label{itemthinstep}
	                             \item  Calculate the residual $\mathbf{r}=\widetilde{\mathbf{p}}-\mathrm{Y}\mathbf{f}_{\rm{thin}}-\mathrm{Y}\mathbf{f}_{\rm{thick}}$
	                            \item  Calculate the gradient  $\mathbf{d}=\mathrm{Y}^\ast \mathbf{r}$
	                             \item  Update $ \mathbf{f}_{\rm{thin}}=\mathbf{f}_{\rm{thin}}+\mathbf{d}$
	                              \item Soft threshold ${\rm{Re}}(\mathbf{f}_{\rm{thin}})$ and ${\rm{Im}}(\mathbf{f}_{\rm{thin}})$, respectively. Set any values below $\tau$ to zero and update $\mathbf{f}_{\rm{thin}}$
	                              \end{enumerate}

                       \item  Reconstructing the Faraday thick sources

                                      \begin{enumerate} \label{itemthickstep}
                                         \item  Calculate the residual $\mathbf{r}=\widetilde{\mathbf{p}}-\mathrm{Y}\mathbf{f}_{\rm{thin}}-\mathrm{Y}\mathbf{f}_{\rm{thick}}$
                                         \item  Calculate the gradient  $\mathbf{d}=\mathrm{Y}^\ast \mathbf{r}$
                                         \item  Update $\mathbf{f}_{\rm{thick}}=\mathbf{f}_{\rm{thick}}+\mathbf{d}$
                                         \item  Calculate the wavelet coefficients for both the real part and imagery part of $\mathbf{f}_{\rm{thick}}$, i.e. $\mathrm{W}\cdot{\rm{Re}}(\mathbf{f}_{\rm{thick}})$ and $\mathrm{W}\cdot{\rm{Im}}(\mathbf{f}_{\rm{thick}})$
                     
                                        \item Soft threshold the wavelet coefficients of $\mathrm{W}\cdot{\rm{Re}}(\mathbf{f}_{\rm{thick}})$ and $\mathrm{W}\cdot{\rm{Im}}(\mathbf{f}_{\rm{thick}})$. Set any values below $\tau$ to zero and update $\mathrm{W}\cdot{\rm{Re}}(\mathbf{f}_{\rm{thick}})$ and $\mathrm{W}\cdot{\rm{Im}}(\mathbf{f}_{\rm{thick}})$
                                 
                                         \item  Calculate the inverse wavelet transform for both the real part and imaginary part, respectively, then update $\mathbf{f}_{\rm{thick}}$
                     
                                         \end{enumerate}
                           \item  $\tau=\tau-\delta$
                 \end{enumerate}  
              
 \item Reconstructed Faraday dispersion function $\widetilde{\mathbf{f}}=\mathbf{f}_{\rm{thin}}+\mathbf{f}_{\rm{thick}}$

\end{enumerate}

Note the Faraday thin sources and thick sources are reconstructed separately. This can be helpful when astronomers focus on either Faraday thin sources or thick sources only. 

If we ignore the step \ref{itemthickstep} and set $\mathbf{f}_{\rm{thick}}=0$, the above algorithm will degenerate to CS-RM-Thin. On the contrary, if we ignore the step \ref{itemthinstep} and set $\mathbf{f}_{\rm{thin}}=0$, the algorithm will degenerate to CS-RM-Thick. In this paper, these CS-based rotation measure synthesis methods are implemented in MATLAB. Our code may be found at http://code.google.com/p/csra/downloads\footnotetext[1]{Download the file ``CS\_RM.zip" which includes both CS-RM-Thin and CS-RM-Thick algorithms}.

%

\section{Experimental results}
\label{sec:experiments} 

We have adopted the standard test platform in~\citet{Brentjens:2005p3385}. In this platform there 126 observing channels within Window 1 (0.036 to 0.5m) evenly distributed in $\lambda^2$. Three different Faraday dispersion functions are simulated to test these CS-Based RM synthesis methods for Faraday thin sources, Faraday thick sources and mixed sources cases. 

 
\subsection{Simulation results for Faraday thin sources}
\begin{figure*}
 \begin{center}
   \includegraphics[scale=0.72]{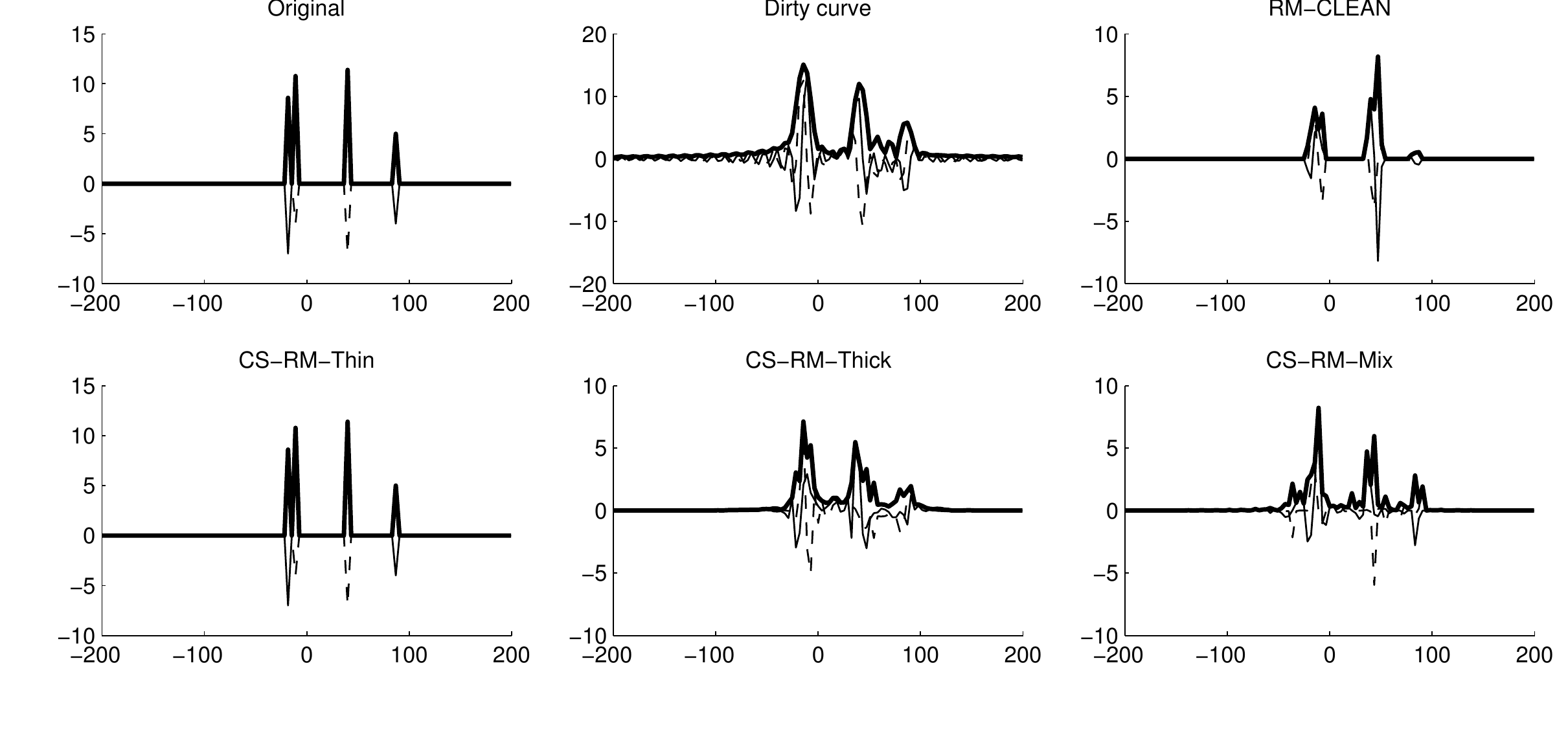}
 \end{center}
\caption{We have tested our methods on a Faraday dispersion function with four Faraday thin sources. From left to right in the first row are:  (a) Original $F(\phi)$, (b) Dirty curve, (c) RM-CLEAN. From left to right in the second row are: (d) CS-RM-Thin, (e) CS-RM-Thick, (f) CS-RM-Mix. The thin solid line shows the real value, the dashed line the imaginary part, and the thick solid line the amplitude. All horizontal axis units are rad~m$^{-2}$, i.e. $\phi$, and all vertical axis units are Jy~m$^2$~rad$^{-1}$.
} \label{fig_rmthin}
\end{figure*}

We simulate a Faraday dispersion function containing four Faraday thin sources. See figure~\ref{fig_rmthin} for the function. From left to right, these sources are: $F(-10)=10-4\mathrm{i}$~Jy~m$^2$~rad$^{-1}$, $F(-17)=-7+5\mathrm{i}$~Jy~m$^2$~rad$^{-1}$, $F(40)=9-7\mathrm{i}$~Jy~m$^2$~rad$^{-1}$ and $F(88)=-4+3\mathrm{i}$~Jy~m$^2$~rad$^{-1}$. The true Faraday dispersion function is shown in the top left corner of figure~\ref{fig_rmthin}. The thin solid line shows the real value, the dashed line the imaginary part, and the thick solid line the amplitude. The Faraday dispersion function in this test is complex valued, i.e. the intrinsic polarization angles are non-zero. We have selected this simulated dispersion function rather than the standard test (real valued Faraday dispersion function) in~\citet{Brentjens:2005p3385} to investigate the behaviour when the intrinsic polarization angles are non-zero. From Eq.~\ref{phiresolution}, we can calculate that the FWHM of the RMSF is around $14$ \rm{rad}~\rm{m}$^2$. The dirty curve which is calculated by assuming all unmeasured emission to be zeros, is shown in figure~\ref{fig_rmthin} (b). The result of RM-CLEAN is shown in figure~\ref{fig_rmthin}~(c). Note that RM-CLEAN cannot correctly reconstruct the magnitude or phase of the Faraday dispersion function in this case. It has been demonstrated previously that RM-CLEAN has difficulty with the separation of sources below the FWHM~\citep{farnsworth2011integrated}. In figure~\ref{fig_rmthin} (a), from left to right, the distance between the first  and the second sources is smaller than the FWHM. In figure~\ref{fig_rmthin} (c), the first source and second source are merged to form an unrealistic source. Results of CS-RM-Thin, CS-RM-Thick and CS-RM-Mix are shown in the figure~\ref{fig_rmthin} (d), (e) and (f), respectively. As one might expect, the result of CS-RM-Thick is poor, because the Faraday dispersion function has no Faraday thick sources. In figure~\ref{fig_rmthin}, though CS-RM-Mix does a better job than CS-RM-Thick and RM-CLEAN in terms of magnitude and phase of the reconstructed disperse functions, the result is far from good enough. The result of 
CS-RM-Thin is shown in figure~\ref{fig_rmthin}) (d). We can see that CS-RM-Thin gives the best result, reconstructing the Faraday dispersion function without any error. This is consistent with CS theory which says that we can reconstruct the sparse signal exactly with ``overwhelming probability"~\citep{Candes:2008p14, Candes:2006p23,Wakin:2008p1623}. In this test, we select $\phi_R=3.6$ \rm{rad}~\rm{m}$^2$ which is around one quarter of FWHM, so $N=480$. 

To carry out a numerical comparison, we use the root mean square (RMS) error to characterise the difference between the reconstructed $\widetilde{\mathbf{f}}$ and the Faraday dispersion function $\mathbf{f}$:
\begin{equation}
\label{equ_rms}
RMS =\sqrt {\frac {\sum_{-N/2+1}^{N/2}(\mathbf{f}-\widetilde{\mathbf{f}})^2}{N }}.
\end{equation} 
The RMS error is calculated for all the candidate methods, and the results are listed in table~\ref{tab:numbericalcoompare}. Results for this test can be found in the first row of the table.  CS-RM-Thin gives the best result.

\begin{figure*}
 \begin{center}
    \includegraphics[scale=0.72]{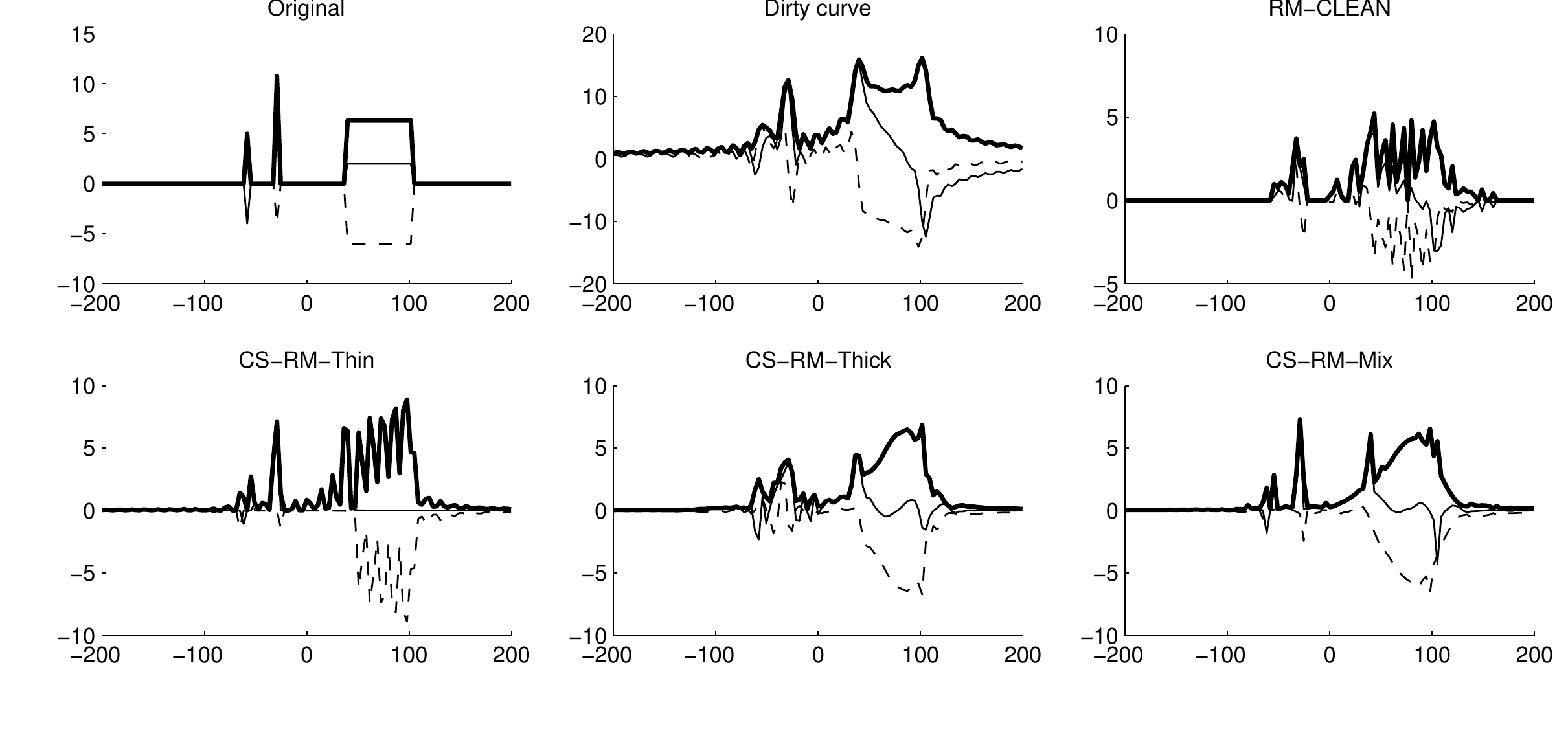}
 \end{center}
\caption{Reconstructed results of a Faraday dispersion function with two Faraday thick sources. From left to right in the first row are:  (a) Original $F(\phi)$, (b) Dirty curve, (c) RM-CLEAN. From left to right in the second row are: (d) CS-RM-Thin, (e) CS-RM-Thick, (f) CS-RM-Mix.} \label{fig_rmthick}
\end{figure*}

 \begin{table*}
\caption{Numerical comparison results (RMS error)}
\begin{center}
\begin{tabular}{ccccccc}
   \hline
            \noalign{\smallskip}
                   & Dirty Faraday dispersion function     & RM-CLEAN     &CS-RM-Thin       &CS-RM-Thick & CS-RM-Mix\\
   \noalign{\smallskip}
            \hline
            \noalign{\smallskip}
               \hline
            \noalign{\smallskip}
Test with Faraday thin sources    & 1.40  &   0.78     & \textbf{0.00} &      0.84	& 0.76\\
Test with Faraday thick sources     & 2.18    &  0.91	 &  1.07 &    \textbf{0.72 }&  0.77\\
Test with Faraday mix sources  & 2.45 &	1.03&	0.95&	0.81&\textbf{0.80} \\
     \noalign{\smallskip}
            \hline
\end{tabular}
\end{center}
\label{tab:numbericalcoompare}
\end{table*}

\subsection{Simulation results for Faraday thick sources}
We now test our CS-based methods for a Faraday dispersion function with Faraday thick sources. Here we assume that the Faraday dispersion function includes two sources $F(\phi)=2-2\mathrm{i}$~Jy~m$^2$~rad$^{-1}$ where $-120\leq  \phi \leq 40$ and $F(\phi)=-6-3\mathrm{i}$~Jy~m$^2$~rad$^{-1}$ where $30\leq  \phi \leq 70$. The simulated Faraday dispersion function is shown in figure~\ref{fig_rmthick} (a). We adopt the previous observing window for this test and the
dirty Faraday dispersion function is shown in figure~\ref{fig_rmthick} (b). Note that this only provides us with the approximate shape of the Faraday dispersion function. As mentioned in~\cite{Frick:2010p3495}, the magnitude of $F(\phi)$ indicates the polarized emission of the region with Faraday depth $\phi$ and its phase defines the intrinsic position angle. For the study of polarized emission of galaxies, the magnitude of $F(\phi)$ is very important, and for the study of orientation of the magnetic field perpendicular to the line of sight, the phase information of $F(\phi)$ is crucial. Unfortunately, Brentjens \& de Bruyn' method can not reconstruct reliable phase information for $F(\phi)$. The cleaned version is shown in figure~\ref{fig_rmthick} (c). RM-CLEAN also failed to reconstruct the phase information. CS-RM-Thin does not work well for this case, because $F(\phi)$ is not sparse. The result of CS-RM-Thin is shown in figure~\ref{fig_rmthick} (d). CS-RM-Mix is also used for this test by assuming that there are both Faraday thin sources and thick sources. The result of CS-RM-Mix is shown in figure~\ref{fig_rmthick} (f). The reconstructed result from CS-RM-Thick is shown in figure~\ref{fig_rmthick} (e). Even though CS-RM-Mix gives a much better result than CS-RM-Thin and RM-CLEAN, the result is not as good as that of CS-RM-Thick. CS-RM-Thick provides the best approximation to the original Faraday dispersion function in terms of both magnitude and phase. This is also supported by the numerical comparison in table~\ref{tab:numbericalcoompare}. From the second row of the table, we can see that CS-RM-Thick gives the smallest RMS error $0.72$ which is slightly better than that of CS-RM-Mix $0.77$.

 \subsection{Simulation results for Faraday mixed sources}
\begin{figure*}
 \begin{center}
    \includegraphics[scale=0.72]{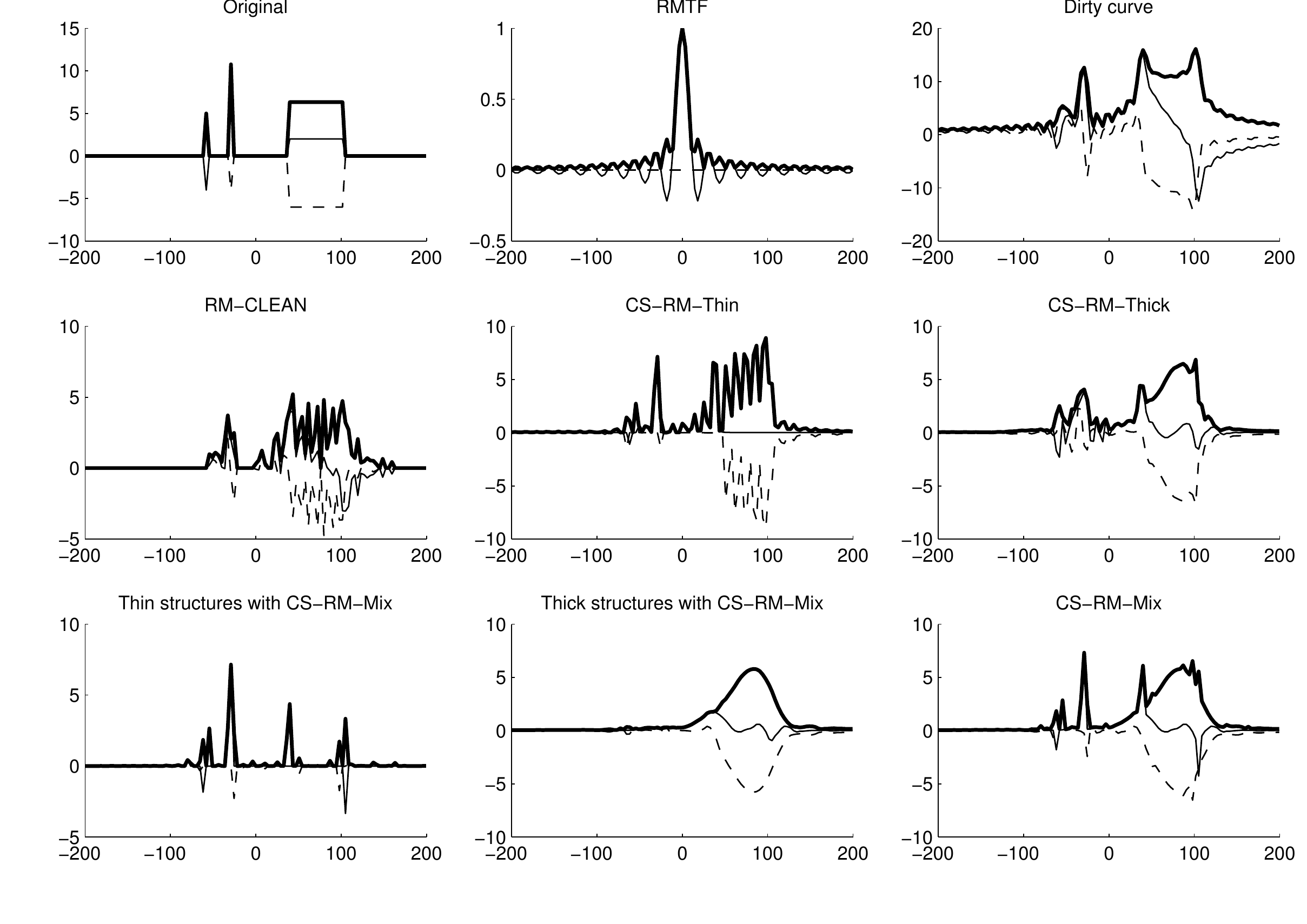}
 \end{center}
\caption{Reconstructed results of a Faraday dispersion function with two Faraday thin sources and a thick source. From left to right in the first row are:  (a) Original $F(\phi)$, (b) RMSF of the observing window with 126 observing channels distributed between 0.036 m to 0.5 m, (c) Dirty curve. From left to right in the second row are:  (d) RM-CLEAN, (e) CS-RM-Thin, (f) CS-RM-Thick. From left to right in the third row are: (g) Thin components $\mathbf{f}_{\rm{thin}}$ by using CS-RM-Mix, (h) Thick components $\mathbf{f}_{\rm{thick}}$ by using CS-RM-Mix, (i) CS-RM-Mix i.e. $\mathbf{f}_{\rm{thin}}+\mathbf{f}_{\rm{thick}}$. All horizontal axis units are rad~m$^{-2}$, and all vertical axis units are Jy~m$^2$~rad$^{-1}$.} \label{fig_rmmix}
\end{figure*}

 So far we have tested our methods for both Faraday thin sources and Faraday thick sources, then we will test our CS-based methods for the mixed circumstances - both Faraday thin sources and thick sources. The simulated Faraday dispersion function for this test is shown in the top left corner of figure~\ref{fig_rmmix}. There are three sources along the line of sight. From left to right, these sources are: $F(-58)=-4+3\mathrm{i}$~Jy~m$^2$~rad$^{-1}$, $F(-30)=10-4\mathrm{i}$~Jy~m$^2$~rad$^{-1}$, $F(\phi)=2-6\mathrm{i}$~Jy~m$^2$~rad$^{-1}$ where $41\leq  \phi \leq 100$. The original Faraday dispersion function is shown in the top left corner of figure~\ref{fig_rmthin}. We use the previous observing window for this test. Based on the observing window (with 126 observing channels distributed between 0.036 m to 0.5 m), the RMSF is calculated and shown in figure~\ref{fig_rmmix}~(b). the dirty curve is shown in figure~\ref{fig_rmmix} (c). The cleaned version by RM-CLEAN is shown in figure~\ref{fig_rmmix} (d). RM-CLEAN performs badly in this test, because there is a Faraday thick source. In general, RM-CLEAN can only work well when there are Faraday thin sources along the line of sight. The results of CS-RM-Thin and CS-RM-Thick are shown in figure~\ref{fig_rmmix} (e) and (f), respectively. We can see that CS-RM-Thin reconstructs the two Faraday thin sources nicely, but failed to reconstruct the Faraday thick source. On the contrary, CS-RM-Thick can properly reconstruct the Faraday thick source, but expends the two Faraday thin sources by mistake. As introduced above, CS-RM-Mix can separate the Faraday thin components $\mathbf{f}_{\rm{thin}}$ and the Faraday thick components $\mathbf{f}_{\rm{thick}}$ during the reconstruction. In this test, the soft-threshold $\tau=1$ and $\delta=0.001$ in the proposed algorithm for CS-RM-Mix. The results of CS-RM-Mix are shown in the last row of figure~\ref{fig_rmmix}. The separated Faraday components $\mathbf{f}_{\rm{thin}}$ and $\mathbf{f}_{\rm{thick}}$ are shown in figure~\ref{fig_rmmix} (g) and (h), respectively. The sum of the separated components is the reconstructed Faraday dispersion function which is shown in figure~\ref{fig_rmmix} (i). We can see that the result of CS-RM-Mix takes advantage of the results of both CS-RM-Thin and CS-RM-Thick, and it gives the closest approximation to the original $F(\phi)$. From objective evaluation point of view, CS-RM-Mix gives the minimum RMS error $0.80$ which can be seen from the third row of table~\ref{tab:numbericalcoompare}.

  \subsection{Discussion}
  
In the figures above, we show the reconstructed dispersion function without smoothing and addition of the residuals - a step commonly know as restoring. For real applications, restoration is an option if the robustness is insufficient. 
  
 From the above three tests, we can see that there is no single CS-based RM synthesis method with an outstanding performance for any circumstances. The best reconstruction can only be achieved when we have some prior knowledge about the Faraday dispersion function and select the relevant CS-based RM synthesis method. If such information is not
 available, CS-RM-Mix can always be used as a compromise. Another option is that we can either select CS-RM-Thin with a large $\phi_R$ or CS-RM-Thick with a small $\phi_R$. If a large $\phi_R$ is selected, $\mathbf{f}$ is likely to be a sparse vector, so CS-RM-Thin should be selected for the reconstruction. On the other hand, a small $\phi_R$ can expand compact sources into extended sources, so CS-RM-Thick will become suitable for the reconstruction. It does not mean that CS-RM-Thick can be used for any cases with a small $\phi_R$, because a smaller $\phi_R$ (which means a larger $N$ from Eq.~\ref{gridnum}) brings more unknowns in $\mathbf{f}$ and more uncertainty. We have to balance these factors.



For RM synthesis, the observing window is quite similar to the frequency filters. For example, the previously introduced observing Window 1 is like a low pass filter in the wavelength squared domain. Therefore, we should bear in mind, to observe radio sources with Faraday thick sources under the same restriction of $m$ and $\delta \lambda^2$, the higher frequency observing band the better.

These CS-based RM synthesis methods are not limited to the optimisation methods L1-Magic solver, FISTA and ISTA. Other L1 norm optimization solvers can also be adopted for solving Eqs.~(\ref{e:faraday_thin}), (\ref{e:faraday_thick}) and  (\ref{e:faraday_mix}). Though the wavelet transform is used as the sparse representations dictionary in this paper, there are some other potential basis functions can be used to achieve sparsity for the Faraday thick sources.   
 
In summary, the performance of CS-based RM synthesis methods depend on the observing window, the resolution of $\phi$, the number of measurements, and the sparsity of the Faraday dispersion function. The reconstruction of these CS-based RM synthesis methods take less time than RM-CLEAN in general. For example, CS-RM-Thin takes 3 seconds for the above tests in contrast with 5 seconds of RM-CLEAN. The calculation time really depends on the construction of the matrix $\mathrm{Y}$, the larger $N$ and $m$ ($N=480$, $m=126$ for the above tests), the more time it takes. The computer is a 2.53-GHz Core 2 Duo MacBook Pro with 4GB RAM.

\section{Conclusions}
\label{sec:conclusion} 

Faraday rotation measure synthesis is a very useful tool to study astrophysical magnetic fields. The problem in RM synthesis is to reconstruct the Faraday dispersion function given incomplete observations. From CS, we know that a signal with sparsity can be well reconstructed based on few measurements. We propose three CS-based RM synthesis methods by finding sparse representations of the Faraday dispersion functions $F(\phi)$ for different circumstances. CS-RM-Thin, CS-RM-Thick and CS-RM-Mix can be used for Faraday thin sources only, thick sources only and mixed sources, respectively.
In general, Faraday thin sources show sparsity in the Faraday depth domain $\phi$, therefore, we apply the CS reconstruction methods (L1 norm optimization solvers) in a straightforward manner i.e. CS-RM-Thin. Although Faraday thick sources are not sparse in the Faraday depth domain, they are sparse in the wavelet domain for a suitably chosen basis wavelet. Therefore, we apply the L1 norm optimization solvers in the wavelet domain i.e. CS-RM-Thick. When there are Faraday mixed sources along the line of sight, we preserve the sparsity by using L1 norm in the Faraday depth domain and the wavelet domain simultaneously i.e. CS-RM-Mix. 

As shown in the experimental results, the performance of these CS-based methods is markedly superior to the traditional RM synthesis methods~~\citep{Brentjens:2005p3385,Heald:2009p3423} in terms of magnitude and angle of the reconstructed Faraday dispersion function. Exemplified by figure~\ref{fig_rmthin}, both Brentjens \& de Bruyn' method and RM-CLEAN do not work well in disentangling two closely spaced sources. In contrast, CS-RM-Thin can separate the sources. 



\section{Acknowledgements}
\label{sec:acknowledgements} 
We thank Jean-Luc Starck for early discussions on Compressive Sampling. In addition, comments, suggestions and derivations made by Dr Brentjens during the review process are very much appreciated.  

  
\bibliographystyle{aa}
 \bibliography{csiro_cs_rm_bibtex}

\end{document}